\documentclass[10pt,sigconf]{acmart}

\usepackage{booktabs}
\usepackage{soul}
\usepackage{xspace}
\usepackage{graphicx}
\usepackage[inline]{enumitem}
\usepackage{ifthen}
\usepackage{etoolbox}
\usepackage{xstring}
\usepackage{tikz}
\usepackage{pifont}
\usepackage{amsmath}
\usepackage{amsfonts}
\usepackage{listings}
\usepackage{color}
\usepackage{tabularx}
\usepackage[normalem]{ulem}
\usepackage{caption}
\usepackage{subcaption}

\definecolor{dkgreen}{rgb}{0,0.6,0}
\definecolor{gray}{rgb}{0.5,0.5,0.5}
\definecolor{mauve}{rgb}{0.58,0,0.82}

\usepackage{ifthen}

\newboolean{showcomments}
\setboolean{showcomments}{true}
\ifthenelse{\boolean{showcomments}}
{ \newcommand{\mynote}[3]{
    \protect\fbox{\bfseries\sffamily\scriptsize#1}
    {\small$\blacktriangleright$\textsf{\emph{\color{#3}{#2}}}$\blacktriangleleft$}}}
{ \newcommand{\mynote}[3]{}}

\newcommand\sysname{NDN-MPS\xspace}
\newcommand\ndnrpc{NSC\xspace}

\newcommand{\etal}{\textit{et al.}\@\xspace}
\newcommand{\eg}{\textit{e.g.}\@\xspace}
\newcommand{\ie}{\textit{i.e.}\@\xspace}

\newcommand\para[1]{\vspace{0.05in} \noindent \textbf{#1.}}
\newcommand\myquestion[1]{\vspace{0.05in} \noindent \textbf{#1?}}

\def\first{({i})\xspace}
\def\second{({ii})\xspace}
\def\third{({iii})\xspace}

\definecolor{verylightgray}{gray}{0.8}

\usepackage{etoolbox}
\usepackage{xstring}
\DeclareListParser{\doslashlist}{/}
\newcounter{ndnNameComponentCounter}%
\newcommand{\name}[1]{{%
  \setcounter{ndnNameComponentCounter}{0}%
  \renewcommand{\do}[1]{{%
    \ifnumgreater{\value{ndnNameComponentCounter}}{0}{\allowbreak/}{}%
    \ifnumodd{\value{ndnNameComponentCounter}}{}{}%
    \detokenize{##1}}%
    \stepcounter{ndnNameComponentCounter}}%
``{\fontfamily{cmtt}\small\selectfont\IfBeginWith{#1}{/}{/}{}\doslashlist{#1}}''%
}}

\usepackage{etoolbox}
\usepackage{xstring}
\newcommand{\namesm}[1]{{%
  \setcounter{ndnNameComponentCounter}{0}%
  \renewcommand{\do}[1]{{%
    \ifnumgreater{\value{ndnNameComponentCounter}}{0}{\allowbreak/}{}%
    \ifnumodd{\value{ndnNameComponentCounter}}{}{}%
    \detokenize{##1}}%
    \stepcounter{ndnNameComponentCounter}}%
``{\fontfamily{cmtt}\tiny\selectfont\IfBeginWith{#1}{/}{/}{}\doslashlist{#1}}''%
}}

\def\cameraReady{} % set to true

\graphicspath{{figures/}}

\setcopyright{none} % No copyright notice required for submissions
\acmConference[Preprint]{Preprint}{XXX}{XXX}
\acmYear{2021}

\begin{document}

\title{Supporting Multiparty Signing over Named Data Networking}

\ifdefined\cameraReady
\author{Zhiyi Zhang}
\affiliation{
    \institution{UCLA}
}
\email{zhiyi@cs.ucla.edu}

\author{Siqi Liu}
\affiliation{%
    \institution{UCLA}
}
\email{tylerliu@g.ucla.edu }

\author{Randy King}
\affiliation{%
    \institution{Operant Networks}
}
\email{randy.king@operantnetworks.com   }

\author{Lixia Zhang}
\affiliation{%
    \institution{UCLA}
}
\email{lixia@cs.ucla.edu}
\else
\author{Anonymous}
\fi

\renewcommand{\shortauthors}{Anonymous Authors}

\begin{abstract}
Modern digitally controlled systems require multiparty authentication and authorization to meet the desired security requirement.
This paper describes the design and development of \sysname, an automated solution to support multiparty signature signing and verification for NDN-enabled applications.
%We first identify the differences between the existing producer-consumer trust model and the new multiparty model, and propose \sysname as the first attempt to allow NDN to support multiparty signature signing and verification for NDN-enabled applications.
%
\sysname suggests several changes and extensions to the existing NDN security solutions.
First, it introduces a new type of trust schema to support signing and verification for multiple signers under complex policies such as threshold schemes.
Second, it extends the NDN signature format to accommodate multisignature schemes such as BLS signature.
Third, it introduces a signature collection protocol to solicit signatures securely from multiple signers.
We further evaluate \sysname by assessing its security properties and measuring its performance.
\end{abstract}

\maketitle

\section{Introduction}
\label{sec:intro}

For reasons such as multiparty contractual policies and mitigating single points of failure, many real-world systems require a joint decision-making process where multiple parties are involved.
For example, for grid-connected distributed energy resources (DER) systems, the conventional way of having single proprietary connections to secure grid assets will no longer be sufficient due to the business model change~\cite{johnson2021recommendations, smart-grid-security}.
Instead, smart devices like solar inverters will have multiple parties, including customers, manufacturers, grid operators, who need to access and send remote commands, often over the public Internet.

Recently, Named Data Networking (NDN)~\cite{ndn2014} started being explored to provide secure networking support for DER systems.
The diversity and number of stakeholders and DER service provider business models~\cite{johnson2021recommendations} require a multiparty trust model with expressive and flexible trust policies.
While NDN has developed supporting mechanisms to secure producer-consumer communications by using crypto signature schemes such as RSA and ECDSA, there is no existing work on multiparty authentication and authorization support.
This paper fills in the gap by designing and implementing a secure, efficient, and usable multiparty signature solution for NDN-enabled applications.

Supporting multiparty signature requires a different security model than that supported by the existing NDN security solutions~\cite{zhang2018security}, which considers only two types of parties: a consumer (verifier) authenticates its received data generated by a producer (signer) based on given policies.
First, multiparty signing involves third-party signers in the system who need to sign data produced by others.
Second, the verifier will need to verify the data's signatures against the given set of signers, so a new type of trust schema is needed to define the set of signers that can jointly make a legitimate signature.
Third, effective and secure coordination among multiple signers is required to ensure the correctness of the joint signing process.

%in each signing process, a coordinator may be needed to take a piece of unsigned data as input and orchestrate the signing process by collecting required signature pieces from a given list of named parties.
%
%In addition, new trust schema policies are needed to help signers to know which entity is allowed to be a coordinator.

\para{Contribution}
To fulfill the gap, this paper proposes \sysname, an NDN-based multisignature signing and verification toolset.
To simplify the orchestration of the signing process among multiple signers, \sysname makes a design choice of having a coordinator in each multiparty signing process who will take the unsigned data as input and carry out the signing process by collecting required signature pieces from all the required signers.
As such, \sysname provides applications with the support of coordinators, signers, and verifiers in multiparty signing scenarios:
\begin{enumerate} [leftmargin=*]
	\item A new design of trust schema that supports the semantics for multiparty signing and verification, including threshold-based policies (\eg, legitimate if signed by any $k$ out of $n$ signers).

	\item A signature collection protocol for the collection of signatures from multiple signers.
	The protocol ensures the authenticity of the collected signatures and preserves the confidentiality of the data being signed and the secrecy of the signing progress.

	\item A new type of key locator scheme to ensure the data integrity in the multiparty signing process and to accommodate multiple signers.
\end{enumerate}
We implement \sysname in C++ with a multisignature scheme called BLS~\cite{bls-paper}\footnote{\sysname can easily be extended to other non-interactive multisignature schemes like MSP multisignature~\cite{boneh2018compact} and the Bitcoin ECDSA threshold signature~\cite{bitcoin-multi-ecdsa, bitcoin-multi-ecdsa2}.}.

\para{Outline}
In the rest of the paper, we first introduce the background and motivation of our work in \S\ref{sec:background}.
We then define the system model of NDN-based multiparty signing and verification in \S\ref{sec:model},
and point out the technical challenges in realizing this model in NDN.
After that, we present the design of \sysname in \S\ref{sec:design} and describe the threshold support in \S\ref{sec:threshold}.
We then assess the security of \sysname in \S\ref{sec:assessment}, evaluate \sysname in \S\ref{sec:evaluation}, and discuss a number of design choices and issues in \S\ref{sec:disc}.
Finally, we conclude our work in \S\ref{sec:conclusion}.

\section{Background and Motivation}
\label{sec:background}

\subsection{Crypto Support for Multiparty Signing}
A conventional method is to collect signatures from individual signers and present the collection to the verifier.
However, this method not only needs a collection of signatures that can be large in size, but also requires the verifier to verify each signature in the collection.
Therefore, space and time complexities grow linearly as the number of signers grows.

A multiparty signature scheme provides an efficient alternative, where a collection of signatures can be aggregated into a single one; the verifier only needs to perform one verification operation against the public keys of all the signers.
A typical example is BLS signature scheme~\cite{BlsSignatureFrc, bls-paper}.
BLS does not require additional negotiation among signers (\ie, plain public-key model), and an arbitrary number of BLS public keys and signatures can be easily aggregated without knowing the private keys.

\subsection{Related Works}

While the concept of multiparty signing~\cite{desmedt1987society} and different multisignature schemes~\cite{harn1994group, shoup2000practical, boldyreva2003threshold, bls-paper} have existed for years,
most of the existing works focus on using the multisignature to combine multiple separate signatures for space and time efficiency, assuming that signatures have already been collected.
In these papers, the joint signature can be viewed as a compressed collection of individual signatures, but not a multisignature on the same message.
For example, Zhao~\etal suggests aggregating signatures to reduce the message length for signing connection path messages in Secure BGP~\cite{sbgp-bls}.
Maxwell~\etal proposes the use of BLS signature in bitcoin or in general blockchain systems to save space in a block~\cite{bitcoin-bls}.
Multisignature has also been used to optimize the permissioned distributed ledger by combining multiple endorsements from different peers into one~\cite{fabricBls}.

In addition to compressing signatures, multisignatures have also been recently used in the digital wallet design in blockchain systems, such as Bitcoin~\cite{bitcoin} and Ethereum~\cite{ethereum}, to increase security by avoiding single points of failure.
The main idea is to split an account secret to store in multiple entities (\eg multiple devices owned by users).
Any transaction from the wallet must be jointly signed by a certain number of secret holders~\cite{bitcoin-multi-ecdsa, bitcoin-multi-ecdsa2}.
However, the trust policies in these systems are largely static (\eg, 1-of-2 account always assumes a 1-out-2 policy), and the signature collection process is usually application-specific and not automated (\eg, copy and paste signature pieces from other devices).

To the best of our knowledge, there is no existing work that supports application-independent multiparty signing with an automated signature collection.  Leveraging NDN's semantically named and secured data concept, our goal is to provide a systematic solution of multiparty signing in NDN with usable supporting components like a signature collection protocol and a flexible trust schema language.

%Therefore, by and large, the existing use of multisignature schemes does not serve the purpose of multiparty trust.
%This is because in these systems, the data covered by each signature is different, so there is no joint trust put from multiple parties.
%
%However, the multisignature on the wallet in blockchain systems such as bitcoin~\cite{bitcoin} and ethereum~\cite{ethereum}.
%Threshold signatures that are compatible for bitcoin is proposed, and it embedded as BLS signature in ethereum~\cite{bitcoin-bls}.
%For these wallets, the bitcoin account secret is distributed to multiple different locations to increase security;
%A certain number of secret pieces are required out of all pieces to initiate a transaction, depends on the construction.
%This approach eliminates a single point of failure, preventing the coins to be stolen or lost when locations are compromised or damaged.
%In addition, this multisignature approach can be used to create joint accounts, so multiple parties can possess a part of the same account~\cite{bitcoin-joint-trading}.
%There are bitcoin client applications that provides multisignature functionalities, for example, Electrum~\cite{electrum-musig}.

%However, these blockchain signatures are intended to provide security for the same identity or only a limited set of identities that knows and trusts each other.

\subsection{A New Trust Model Is Needed}

\para{The Existing Producer-Consumer Trust Model} \\
The producer-consumer trust model is adopted by the existing NDN security support.
Under this model, each NDN data object is assembled, named, and cryptographically signed by its producer.
To help consumers to verify the data, each data object carries a signature information field to keep the meta information of the signature;
the most important information is the key locator, indicating which key should be used to verify the signature value.
A key locator is usually the name of the verification key so that a consumer can fetch the corresponding key to verify the received data.

Verifying the signature against a trustworthy public key only ensures authenticity. To further verify the authorization of the producer, a trust schema is used to determine whether the producer is legitimate to produce the named data.
NDN's semantic naming enables the trust schema to use names to systematically define the relationship between the signing key and data object of each step in an acceptable certificate chain.
Trust schema rules for a specific application system usually contain the following elements:
\first the format or structure of the data names that the rules apply to,
\second the name pattern of the expected signing key name at each step in the certificate chain,
\third the acceptable trust anchors that certificate chains end with.
If a signature is generated by a party whose key name is not allowed for signing such data or the signer's identity cannot be certified by an acceptable certificate chain to a trusted anchor, the signature will be rejected.

Importantly, the producer-consumer trust model only \emph{assumes two parties}, the producer and the consumer, both of which are directly involved with the data.

\para{The Need for a New Model}
In a multiparty signing scenario, the existing trust scheme is insufficient to clearly describe the policies of the new trust model
because
\first there are multiple signers,
\second (at least some) signers are not the producer of the data to be signed; and
\third multisignature requires a new party that coordinates the signing, which the trust schema should clearly define.
The current protocol specification and implementation of NDN libraries also need enhancement to support multiparty signing.
For example, the current key locator cannot carry sufficient meta information of multiple signers, and the trust schema does not support the signing and verification of multisignatures.

\subsection{An Example Scenario}
\label{subsec:example}

We introduce an example scenario used throughout the paper to ease the reader's comprehension of various concepts.
Assume an equipment manufacturer requests customers upgrade an inverter’s firmware to correct a security vulnerability.
To make the update, the maintenance operator, Alice, needs a command, an NDN data object, to be jointly signed by the following parties:
\first the equipment manufacturer’s Quality Assurance (QA) department (certifying the upgrade is of release quality), \second the solar site’s operation team (ensuring they know the equipment will be offline for some time), and \third the site’s owner (permitting the possible production reduction, if the process cannot be completed promptly).
The equipment that executes the firmware update will verify the signed data to ensure that it can safely proceed with the update.
We further assume Alice's NDN name is \name{/Site/maintenance/Alice}, the signers' prefixes are \name{/Mfr/QA}, \name{/Site/operation/}, and \name{/Site/Owner}, respectively.
Under each prefix, there can be multiple entities; for example, under \name{/Mfr/QA} there can be \name{/Mfr/QA/operatorX} and \name{/Mfr/QA/operatorY}.

\begin{figure}
\centering
\includegraphics[width=0.9\linewidth]{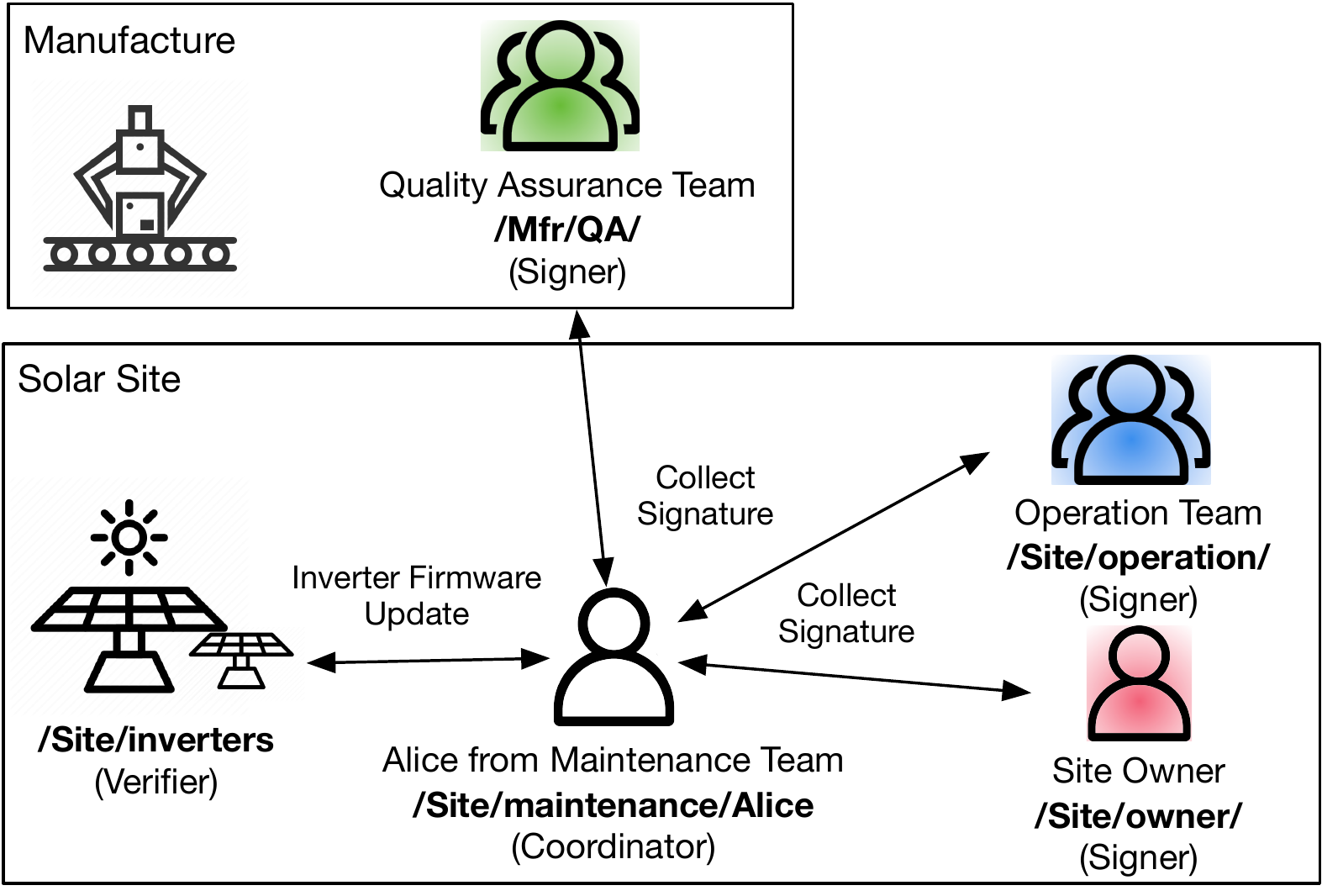}
\caption{Example Scenario}
\vspace{-2mm}
\label{fig:example}
\end{figure}

\section{Models And Challenges of NDN Multiparty Signing}
\label{sec:model}

The goal of \sysname is to support multiparty signature signing and verification over NDN.

\subsection{System Model}
\label{sec:model:model}

In each multiparty signing process, there are three parties:
\begin{itemize} [leftmargin=*]

	\item \textbf{Coordinator}.
	The coordinator is signaled by the data producer to generate a multisignature for the data.
	Based on the application's requirements defined by the trust schema, the coordinator will collect signatures from a group of known signers and then aggregate them into a multiparty signature.
	To collect signatures, the coordinator needs to know signers' names and certificates.

	\item \textbf{Signers}.
	The signers can come from different organizations and have different trust anchors.
%	Multiple signers need to jointly generate a multisignature.
	If a signer approves the data to be signed, either through in-band operations (\eg, database query) or out-of-band operations (\eg, human decisions), that individual will make a signature piece for aggregating with other signers' pieces.

	\item \textbf{Verifier}.
	The verifier is called by the data consumer to verify a multisignature along with the data covered by the signature.
	The verification succeeds if the multisignature is valid and if its signers satisfy the application's requirements as defined in the trust schema.

\end{itemize}
It is noteworthy that we separate the above roles based on functional logic.
In practice, one entity can play multiple roles, \eg, the coordinator can also be one of the signers.
In addition, an entity can play different roles in different signing processes.

In our example (\S\ref{subsec:example}), the operator Alice is the producer of the command and runs the coordinator.
The manufacturer's QA team, the site's operation team, and the site owner are the signers.
After being signaled by Alice that the signed command is ready,  the inverter, as the data consumer, will fetch the data and call the verifier to verify the multisignature.

\para{Assumptions}
We assume all three types of entities have gone through the standard NDN bootstrapping process~\cite{zhang2018security}, through which each entity in the system obtains an identity, certificate, the system trust anchors, and the trust schema.
Note that all the signers' key types should be compatible with \sysname, \eg, BLS key type with the current \sysname implementation.
Therefore, signers can authenticate coordinators, and verifiers can authenticate signers.
As stated, we also assume the coordinator's awareness of the signers from which the coordinator can select appropriate signers to generate legitimate signatures.
In addition, there is a known attack against the BLS signature, called rogue key attack~\cite{BlsSignatureFrc}, where a malicious coordinator may successfully generate a legitimate multisignature without obtaining real signature pieces from the signers.
To prevent such types of attacks, as suggested by \cite{BlsSignatureFrc}, we assume that each signer's certificate carries a piece of cryptographic information, called a proof of possession, to prove the possession for the public key corresponding to secret key.

%\todo{rephrase: standard + special pop}
%Before a coordinator starts a new signing process, we make the following assumptions about the system.
%\begin{itemize} [leftmargin=*]
%	\item \todo{not needed}. Reachable prefix. The coordinator $I$ knows the reachable prefix of each signer $S_i$ in the network.
%	If a signer's prefix is not publicly reachable, the coordinator will also need a forwarding hint to help reach the signer.
%
%	\item Public keys installation.
%	The verifier $V$ has installed the public key along with the identity name of each signer.
%	Such information can be obtained in a form of an NDN certificate with inline or out-of-band means.
%
%	\item Proof of possession. When installing a public key of a signer, a proof of possession (PoP) or equivalent (\eg a self-signed certificate) should be provided by the signer to prevent rogue key attack~\cite{BlsSignatureFrc}.
%		  \siqi{This is special, so emphasize it}.
%	\item Awareness of the trust schema.
%	The coordinator $I$ is aware of the trust schema to generate a sufficient signature for verifier $V$ to validate.
%	In addition, each signer has the trust schema for validating the identity of the coordinator.
%\end{itemize}

\subsection{The Signing and Verification Procedure}
\label{sec:model:procedure}

Under the system model, a complete signing and verification process can be described as follows in an abstract way.

\begin{enumerate} [leftmargin=*]
	\item The coordinator starts by deciding the signer list based on the trust schema and proceeds through the signing process by contacting individual signers.

	\item Each signer verifies the request against the trust schema to ensure that the coordinator is legitimate.

	\item If the signer is permitted by the application logic to sign the data, it generates the signature and sends it to the coordinator.

	\item The coordinator aggregates the collected signature pieces into one and appends it to the unsigned data to get the finalized signed data object.

	\item After the consumer application obtains the signed data, the verifier verifies the signed data against the trust schema.
\end{enumerate}

The coordinator is triggered to start the process by certain application events.
For example, the maintenance operator Alice can start the coordinator when receiving a firmware update notification from the manufacturer.
Similarly, obtaining the signed packet is also realized at the application layer.
In the same example, the inverter can call the verifier when receiving the firmware update command.

To ensure the security of the system, the following policies should be explicitly and rigorously defined by the trust schema.
\begin{itemize} [leftmargin=*]
	\item \textbf{Coordinator Verification Trust Schema(Section~\ref{sec:model:verification-schema})}.
	In step 2, the trust schema needs to define the coordinator based on the names or name patterns of legitimate coordinators.

	\item \textbf{Multisignature Trust Schema(Section~\ref{sec:model:schema})}.
	In step 5, the trust schema needs to specify how many and which signers can generate a legitimate signature for a given named data.
\end{itemize}

In step 3, whether a signer will agree to the sign data depends on two factors.
First, each signer should ensure the data is allowed to be signed by her according to the installed multisignature trust schema.
Second, the signer may also need approval from the application logic, \eg, through some database query or even human decision making; this process is application-specific and thus omitted in this paper.
Importantly, depending on application scenarios, the time duration of this step may vary significantly.
In the example in Section~\ref{subsec:example}, to ensure the power supply after shutting down the inverters for the update, the public utility may need to measure the remaining power supplies against the power demand in the area before agreeing to sign the command.

\subsection{Coordinator Verification Trust Schema}
\label{sec:model:verification-schema}

The coordinator verification schema is to verify the identity and authorization of the coordinator when receiving a signature request.
Such a request carries a regular signature generated by the coordinator, and thus this trust schema can be expressed using the existing trust schema languages or tools~\cite{nichols2019lessons, cxxschema}.
In our example (\S\ref{subsec:example}), the trust schema can define only maintenance operators, who are under the prefix of \name{/Site/maintenance} and derived from the anchor certificate \name{/Site/maintenance/KEY/123}, can be coordinators sending signature collection requests.
The pseudo-code of such a trust schema policy is shown below.

\begin{figure}[h]
\begin{lstlisting}[language=TeX]
Signature Request Name:
  /<SignerPrefix>/MPS/request/Site/maintainence/<operator>/<>
Key Name: /Site/maintenance/<operator>/KEY/<>
Anchor: /Site/maintenance/KEY/123
\end{lstlisting}
\vspace{-5mm}
%\caption{Coordinator Verification Trust Schema Example}
%\label{code:schema-0}
\end{figure}

\subsection{Multisignature Trust Schema}
\label{sec:model:schema}
The multisignature trust schema requires additional semantics for legitimacy verification of multiple signers for a signature.
In particular, a multisignature trust schema should contain three pieces of information.

\begin{itemize} [leftmargin=*]
	\item \textbf{Data Profile}.
	A profile specifies which data the policy should be applied to. This profile is usually the name or the name pattern of the data.

	\item \textbf{Legitimate Signer List}.
	A legitimate signer list specifies all the signers required to make a legitimate signature for the given data.
	This list can be effectively expressed by a list of the signers' name patterns.

	\item \textbf{Known Signers}.
	As stated, our model assumes that the coordinator knows the signers in the system.
	One implementation option is to include all signer information, including a routable prefix and the certificate for each signer, as part of the multisignature trust schema.
	The known signers can also be imported signers as a table or obtained from trust anchors.

%
%	Trust Anchors provides trust information of individual signers through certificates signed by them.
%	These certificate specifies the individual parties authorized to sign part of the multisignature.
%	To verify the signature of each of these parties, Signer Names and Public Key bits are also provided in the certificates.

\end{itemize}

In the example described in \S\ref{subsec:example}, the data profile should match the name of the firmware update command data, the legitimate signers are the prefixes of three required parties, and the known signers are all the candidate signers in the system.
The pseudo-code of such a trust schema policy is in Section~\ref{sec:design:schema}.

In \S\ref{sec:threshold}, we further extend the syntax of multisignature to support threshold-based policy.
For example, a threshold-based policy can express the legitimate signers as any $k$ out of $n$ candidate signers.

%This includes signers that are listed as trust anchors and those who can be trusted by verifying their certificates against the trust anchors with trust schema.

%The core information is the acceptable signer combinations because they indicate which signer combinations are legitimate.
%For example, acceptable signer combinations can be represented by $S_1 \land (S_2 \lor S_3)$.
%This means $S_1$ and one of $S_2$ and $S_3$ can generate a legitimate signature.

%\subsection{Missing Pieces in NDN Security}
%\lz{You've said this multiple times, I suggest to remove this subsection}
%
%\para{Multisignature Trust Schema}
%As stated, the existing design~\cite{trust-schema} of trust schema assume digital signatures generated by a single signer and thus cannot accommodate the policy in which multiple signers are involved.
%
%\para{Multisignature Signature Information}
%The current signature information field in NDN Data packets assumes only one signer for each packet and thus insufficient to carry multiple signers' information.
%
%\para{Multiparty Signing Support}
%There is a lack of signature collection protocol for a coordinator to provide the unsigned Data and collect signatures from multiple signers.

\subsection{Technical Challenges}
\label{sec:model:challenges}

\para{Signature Coverage in Multiparty Signing}
To be aggregated correctly, each signer's signature should cover identical data.
However, by the current design of the signing process, at the time of signature generation, the signer will customize the signature information file of the data packet by writing the signing key name into the key locator field.
Consequently, each signer will sign the data with customized signature information, thus leading to inconsistent data signed by different signers and a broken final signature after an aggregation.

One potential solution is to let the coordinator hardcode the signature information, and each signer generates a signature without modifying the signature information.
However, this solution is insufficient because the coordinator does not know the final signer list before contacting the signers.
For example, to meet the requirement of a signer under the prefix \name{/Mfr/QA}, the coordinator Alice decides to contact \name{/Mfr/QA/operatorX} and thus hardcode this name into the key locator field.
However, operator X may be offline, and thus Alice needs to find an alternative signer, say \name{/Mfr/QA/operatorY}.
As such, Alice needs to change the signature information halfway through the signature collection.

As described in \S\ref{sec:design}, in \sysname, we propose using a late-binding key locator to address this challenge.

\para{Threshold Signature}
Threshold signature scheme \cite{harn1994group} is a type of multiparty signature, which supports a flexible access structure:
instead of explicitly defining the signer combinations for a legal signature, $k$ arbitrary signers out of a signer set of size $n$ can jointly generate a legal signature.
A common approach to support the threshold scheme is to directly utilize threshold signature schemes.
For example, a typical construction of threshold signature schemes is to combine traditional signature schemes with Linear Secret Sharing~\cite{secret-sharing}.
However, this solution suffers from a few disadvantages.
First, to bootstrap the system, either \first a centralized trusted dealer is needed for key distribution or \second the signers need to exchange a large number of messages for setting up their keys in a distributed manner.
In addition, the key maintenance is complicated because when a key is compromised, or when there is a new signer, it may require a new bootstrapping process where all signers must be online and update their keys.
Lastly, the key aggregation is relatively slow, which takes $O(k^2)$ traditionally and $O(k \log{k})$~\cite{fast-bls-aggregation} to perform the interpolation to construct the secret.

As described in \S\ref{sec:threshold}, \sysname takes a system approach with the multisignature trust schema, bypassing the aforementioned issues.

%
%\todo{no one can understand the problem.}
%\para{Potential Interference to Application Use}
%In the signature collection process, if intermediate results, including the unsigned Data packet and Data packet signed by single signer, are presented in the network, they can potentially interfere the later application operations because these packets carry the same name as the final Data packet carrying the aggregated signature.
%For example, if the verifier sends an Interest to fetch the final packet, this Interest may fetch the unsigned packet or a packet signed by a single signer from the network cache (\ie cache poisoning).
%
%\para{Timing Consideration}
%The signers may need a manual verification or a length computation before agreeing the content of the unsigned data.
%For automatic verification, the turnaround time is usually less than a second; but the manual verification, the turnaround time may be minutes, even hours.
%This time may be longer than the lifetime of the Interest, resulting in a timeout for the coordinator's interest.
\section{Design of \sysname}
\label{sec:design}

\begin{figure}[t]
	\centering
	\includegraphics[width=0.5\textwidth]{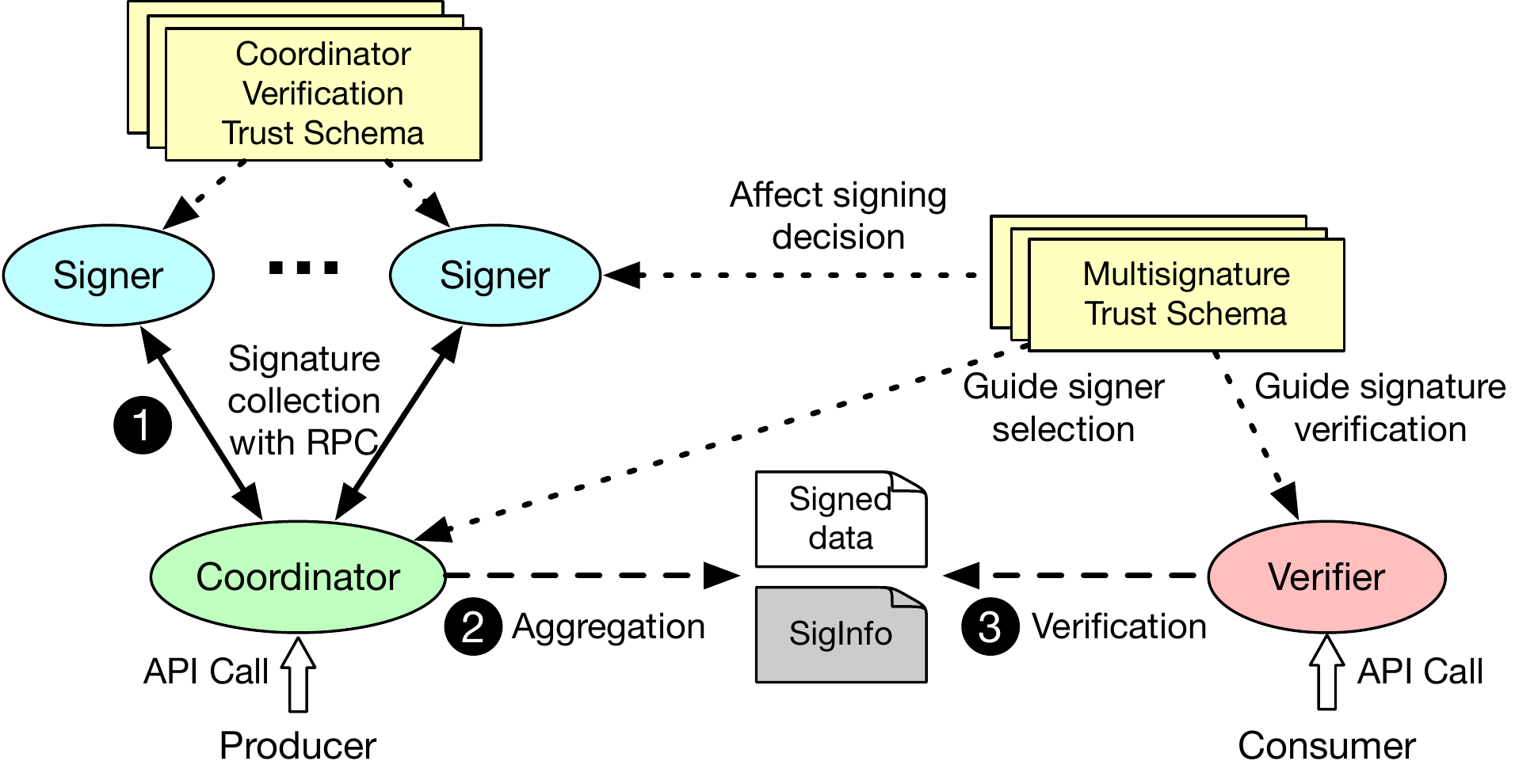}
	\caption{An Overview of \sysname}
	\label{fig:overview}
\end{figure}

We describe the design of \sysname by first giving an overview and then introducing each main component of the \sysname.
This section focuses on how to support multiparty signing and verification when the legitimate signers or their name patterns are specified.
We present the threshold signature support in \S\ref{sec:threshold}.

\subsection{An Overview}
\label{sec:design:overview}

\sysname is designed as a security support toolset for applications.
\sysname consists of three main components (Figure~\ref{fig:overview}).
First, a new type of trust schema that allows applications to express the signing and verification policies of multisignatures.
Second, an extension of the key locator in NDN data objects to keep the to-be-signed data consistent across signers and allow complex signature information.
Third, a signature collection protocol for the coordinator to collect signatures from multiple signers.

These three components are used in the process of multisignature signing and verification.

%\para{Multiparty Signing Process Initiation}
%
%After that, the coordinator will first work out a list of signers whose aggregate signature can satisfy the trust schema (Section~\ref{sec:design:selection}).

\para{Signature Collection}
The coordinator starts the signing process when the application provides the unsigned data and the multisignature trust schema (Section~\ref{sec:model:schema}) through a local API call.
From the trust schema, the coordinator identifies a list of signers and then starts the signature collection.
In a nutshell, the coordinator issues a signed request to the signer and presents the unsigned data as the request parameter.
To ensure the consistency of the data to be signed among signers, the coordinator will fill the key locator field with a placeholder name called late-binding key locator (Section \ref{sec:late-binding}).
Such a placeholder is the key to make the data to be signed deterministic and consistent across multiple signers.

Upon receiving the request, each signer will first verify it against the coordinator verification trust schema to ensure that its sender is a legitimate coordinator.
After that, the signer can decide on whether to sign the data or not.
In this process, besides checking against known multisignature trust schema policies, the signer may also need to go through some application-specific procedures.
If the signer's application agrees to sign the data, it will generate the signature over the data, without modifying the key locator or any other fields.

In an unstable network or in an attack, signers may be forced offline, unable to be reached by the coordinator.
In this case, the coordinator can try to reach an alternative signer permitted by the trust schema, \eg, the coordinator Alice can try \name{/Mfr/QA/operatorY} when \name{/Mfr/QA/operatorX} is unavailable.
This design provides a certain degree of resiliency that even if some signer fails, the signing can continue.
Importantly, the change of signers will not affect the value of the late-binding key locator name as it is a placeholder, so the consistency of the data being signed among signers is preserved.

\para{Signature Aggregation}
After collecting sufficient signature pieces, the coordinator aggregates them to generate a single signature and attach it to the original unsigned data.
In addition, the coordinator generates another Data packet, called a SigInfo packet, to carry carries the signer information, \ie, signers' key names.
Importantly, SigInfo is named with the placeholder key locator name.
As such, a SigInfo packet can be linked from the signed data's key locator, overcoming the limitation of the existing key locator design, which does not support multiple signers.

\para{Signature Verification}
To verify a piece of multiparty signed data, the verifier first needs to obtain both the signed data, which is expected to be integrated with specific applications.
The SigInfo packet can then be fetched with an Interest carrying the key locator name specified in the signed data.
After that, the verifier first checks the signer list from the SigInfo packet against the multisignature trust schema to ensure the signers are legitimate to sign such a piece of data.
Finally, the verifier validates the multisignature using the public keys of the signers.

\subsection{Multisignature Trust Schema}
\label{sec:design:schema}

%We want the trust schema to satisfies most needs for multiparty signature but still remain simple and efficient.\lz{how do we even know what may be the most needs at this time??}
%Specifically, our implemented trust schema that support the following semantics.

As the baseline, the trust schema allows specifying multiple required signers.
For example, a signature is valid only when required signers $R_1$ to $R_n$ are present.
With our trust schema syntax, the key name of $R_1$ to $R_n$ will be listed under a configuration field called ``all-of''.

Furthermore, to enjoy the flexibility of structural names, the trust schema should allow name patterns.
For this purpose, the multiparty trust schema supports wildcard character $*$, which can match any name component.
Using the example described in \S\ref{subsec:example}, the wildcard character in \name{/Mfr/QA/*/KEY/*} allows any operator under the equipment manufacturer's QA team to participate in the joint signing.
%In circumstances where the trust schema requires multiple signers that are of the same name pattern, we also supported an optional ``n $\times$ '' prefix to indicates the required number of matched signers.
%For example, if the system requires two operators from the manufacturer's QA team, namely, \name{2x/Mfr/QA/*/KEY/*}.

The pseudo-code of a multisignature trust schema for the example use case is shown below.
The rule applies to the command to update the solar inverter firmware.
The known signers should be the registered operators under each responsible organization’s prefixes, which are omitted for simplicity.

\begin{figure}[h]

\begin{lstlisting}[language=TeX]
Data profile: /Site/inverters/firmware/update
All-of {     /Mfr/QA*/KEY/*
             /Site/operation/*/KEY/*
             /Site/Owner/*/KEY/*      }
\end{lstlisting}
\vspace{-5mm}
%\caption{Multisignature Trust Schema Example}
%\label{code:schema-1}
\end{figure}

\subsection{Late-binding Key Locator Name}
\label{sec:late-binding}

At the beginning of a signing process, the coordinator does not know the exact final signers out of the eligible list, because some signers may be unavailable.
Therefore, the coordinator may have to change the signer list in the middle of the signing process.
However, before a signer generates a signature piece, the signature information, including the key locator, needs to be deterministic and consistent among all signers so that signature pieces can be aggregated.

Therefore, \sysname lets the coordinator place a placeholder Data packet name as the key locator.
This name is late-binding because, at the time of the signature collection phase, there is no Data packet bound to this name.
After completing the signing process, the SigInfo packet containing the signature information will be available under that name to specify the exact signers of this signature packet.
This packet typically contains the list of the signer's key names; hints such as the trust schema name or the aggregated public key can also be included. 
The SigInfo packet can be signed by the coordinator for data integrity.
Since the exchange of signed data eventually occurs between the coordinator and the verifier at the application layer, the verifier will have the means to authenticate the coordinator and the SigInfo packet.

When verifying an aggregated signature, the verifier will also need the SigInfo packet.
% Different applications can have different way for a verifier to obtain the SigInfo packet.
The SigInfo packet can be fetched by the verifier with an Interest carrying the key locator name.
With the signer information from the SigInfo packet, the verifier can aggregate the signers' public keys accordingly and verify the signature value.

\begin{figure}[t]
	\centering
	\includegraphics[width=0.49\textwidth]{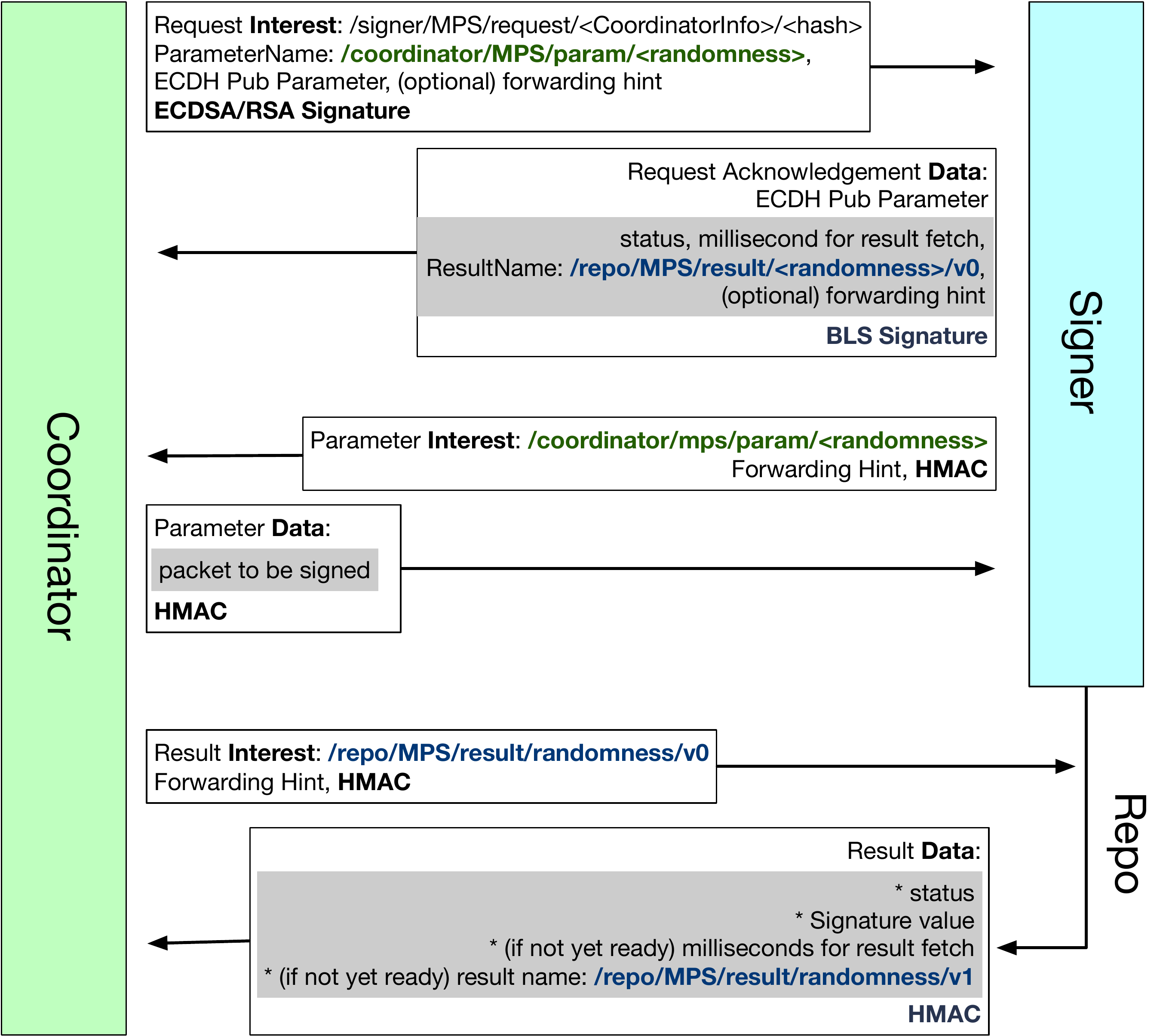}
	\caption{Remote procedure call in \sysname}
	\label{fig:rpc}
\end{figure}

\subsection{\sysname Remote Procedure Call}
\label{sec:ndnrpc}

An important component of multiparty signature support is a protocol to request and collect signature pieces from signers.
This interaction can be considered as a remote procedure call (RPC) where the coordinator is the caller and where the signer is the executor.
The RPC protocol used in \sysname is based on \ndnrpc but provides stronger security and privacy.

\para{Security Targets}
We want to provide authenticity and integrity of the process to prevent attackers from hindering the signing process or voiding the signature by inserting unexpected signature pieces.
Furthermore, the confidentiality of the process is also important, considering attackers may increase their advantages by obtaining useful information before the signed data is finalized.
In addition, potential record-and-replay attacks and denial-of-service attacks should be considered as well.
We assess the security of \sysname's RPC in \S\ref{sec:assessment}.

\para{Related Works}
A straightforward solution is to directly apply existing NDN-based RPC protocols between the coordinator and each signer, for example, RICE~\cite{krol2018rice}, the RPC used in DNMP~\cite{nichols2019lessons}, and Named Service Call (\ndnrpc)~\cite{ndnnsc}.
However, while RICE provides sufficient security and privacy protection, it requires modification of the underlying NDN forwarding pipeline.
DNMP RPC requires an NDN synchronization protocol which is excessive for our use case.
In comparison, \ndnrpc is lightweight and requires no modification of the NDN forwarding, but it does not provide payload encryption and result confidentiality as needed by \sysname.
Therefore, in this work, we extend the design of \ndnrpc to fit our needs.

\para{\sysname RPC}
We design the \sysname RPC for signature collection based on \ndnrpc but provides stronger security and privacy.
The main design choice made by \sysname's RPC is the following.
\begin{itemize}[leftmargin=*]
	\item Using signed Interest and Data exchange for authenticity and integrity.
	Coordinator trust schema is also applied when signers verify the RPC request from coordinators.

	\item Applying Diffie-Hellman (DH) key agreement to negotiate a per-RPC shared secret for packet payload encryption.
	As such, all the sensitive information exchanged in the RPC is encrypted for privacy protection.

	\item Wrapping the parameter and the result of RPC into a Data packet whose name suffix is randomly generated and content is encrypted, as shown in Figure~\ref{fig:rpc}.
\end{itemize}
Through this design, attackers cannot know what data is signed in the signature collection and have no idea about the result of each RPC, which satisfies one of the main design goals of \sysname.

As shown in Figure~\ref{fig:rpc}, to start the signing process, the coordinator expresses a signed request Interest to the signer.
This Interest carries a name, called $ParaName$, pointing to a parameter Data packet carrying the unsigned data.
At this time, the parameter Data packet is not generated yet.
Such a name comprises the coordinator's prefix and a randomly generated component, revealing no information about the unsigned packet.
If the coordinator's prefix is routable, to make the parameter Data packet available, a forwarding hint to the packet should be carried by the request as well.
In addition, the Interest also carries DH public parameters.

On receiving the request Interest, the signer will first verify the signature to ensure the coordinator is a legitimate party according to the coordinator verification trust schema defined by the system.
Then, the signer will also generate a pair of public and private keys for DH and derive a message authentication code key $K_{mac}$ and a symmetric encryption key $K_{enc}$ from the DH (key derivation function is used along with DH for better security).
After that, the signer will generate an acknowledgment Data packet containing the following information.
\first A status code indicating whether the request is accepted. Since the signer has not seen the data to be signed, the status solely relies on the coordinator identity verification result.
\second The estimated time for the RPC result to be available.
\third The name of the result Data packet, called $ResultName$. This name is supposed to be random and unrelated to the signer.
For this purpose, the final result Data can be published to a third-party repository, and a forwarding hint to that repository can be added to the acknowledgment Data as well.
Importantly, these parameters are encrypted with the $K_{enc}$.
In addition, the acknowledgment Data also contains the signer's public parameters in plaintext for DH.
This acknowledgment packet is signed by the signer's private key.

After replying with the acknowledgment, the signer can immediately issue an Interest to fetch the parameter.

Upon receiving the acknowledgment packet, the coordinator verifies the signature, calculates the same $K_{mac}$ and $K_{enc}$ from DH with the coordinator's public key, and decrypts the payload carried in the Data content.
Then, the coordinator will finalize the parameter packet by naming it as $ParaName$, encrypting the unsigned packet with $K_{enc}$ into the content field of the Data, and signing it with $K_{mac}$.
This parameter packet will be fetched by the signer's Interest packet.
After knowing the $ResultName$ and time for fetching the result from the acknowledgment packet, the coordinator can schedule a data fetch accordingly.

On the signer side, after fetching the parameter packet, if the signer agrees to sign the packet, the signer will generate a signature over the unsigned packet.
During the signing process, the signer will not modify any field, including the signature information field, of the unsigned packet.
Then the signer will encrypt the signature value with $K_{enc}$ and wrap the ciphertext into a Data packet named as $ParaName$ and signed with $K_{mac}$.

%\para{Use of Forwarding Hint}
%In the signature collection process, the signer will need to fetch the parameter packet and the coordinator will need to fetch the result packet.
%For the parameter packet fetch, if the coordinator's name prefix may not be reachable through the Internet, a forwarding hint to the coordinator's network can be inserted into the request Interest.
%For the result fetch, the forwarding hint is used for the purpose of privacy instead of reachability.
%As stated, to prevent an attacker link the coordinator's request and the signer's result, the signer can randomly name the result and insert it to a network repository.
%Since a random name prefix is not reachable, the forwarding hint can be used to guide the result fetching Interest into the repository.

%\para{Security and Privacy}
%In the signing process, all the sensitive content after the first Interest packet of the RPC is encrypted for privacy protection.
%Even in the first Interest packet, the parameter name leaks no information of the packet to be signed and the parameter packet's content is encrypted.
%Since $Key_{mac}$ and $Key_{enc}$ are derived with DH per RPC, \sysname provides forward secrecy.
\section{Support of Threshold}
\label{sec:threshold}

\sysname provides the function of threshold signature by a system means with the multisignature trust schema.
This departs from the conventional solution of applying additional cryptographic tools (\eg, linear secret sharing~\cite{secret-sharing}) to existing signature schemes.
As stated in \S\ref{sec:model:challenges}, this is because the cryptographic solution suffers from several drawbacks including high maintenance overhead and low efficiency.

Specifically, we let multisignature trust schema support threshold verification policy by adding two new field called ``at-least-num'' and ``from''.
To express the policy that a signature is valid when $k$ out of $m$ signers are present, the $k$ will be specified by the ``at-least-num'' and the key names of $m$ signers are listed after the ``from''.
We show an example of multisignature trust schema with threshold verification below.
Different from the required signers as described in \S\ref{subsec:example}, it only needs any two of the three parties to generate a legitimate signature.
Note that these two fields can also be used together with existing multisignature fields to support more complex policy, \eg, a signature needs $n$ requires signers and $k$ out of $m$ optional signers.

\begin{figure}[h]
\vspace{-3mm}
\begin{lstlisting}[language=TeX]
Data profile: /example/data/*/*
At-least-num 2
From {      /Site/operation/*/KEY/*
            /Mfr/QA/*/KEY/*
            /Site/Owner/*/KEY/*     }
\end{lstlisting}
\vspace{-5mm}
%\caption{Multiparty Trust Schema Example}
\label{code:schema}
\end{figure}

Compared with cryptographic solutions of threshold scheme, \sysname does not need any centralized trusted dealer, and the key maintenance is much simpler.
To be more specific, the key setup does not require interactive operations and can be done individually per signer with the public parameters.
In addition, the keys can be managed individually.
For example, a signer's key can be renewed or revoked without bothering other signers' keys.

\section{Security Assessment}
\label{sec:assessment}

%\begin{itemize}
%	\item Authentication: correctness of signature
%	\item Privacy: no information leak in the negotiation protocol: request ID, signature are all random, DH + AES encryption to hide the data to be signed.
%\end{itemize}

We start the security analysis of \sysname against the desired properties as stated in \S\ref{sec:ndnrpc}
Then, we consider several attacks potentially conducted by attackers.

\para{Multiparty Authentication}
The verifier can authenticate the signers of the aggregate signature and judge whether the signature is legitimate by the features of the underlying BLS signature and the multisignature trust schema.
The signature verification will guarantee the following:
\begin{itemize}[leftmargin=*]

\item The signers are qualified to generate a legitimate signature required by the multisignature trust schema.

\item The signature value can be verified by public keys of claimed signers. This protection is ensured by the BLS signature that can only be verified with corresponding signers' keys.

\item The data content is not altered. If data has changed after the signature pieces are created, the BLS signature will no longer be valid for the altered data.
\end{itemize}
If signature verification succeeds, this ensures the authentication of all involved signers.

\para{Coordinator and Signer Authentication}
In \sysname, any signer can verify the identity of the coordinator by verifying the signature carried by the RPC request Interest.
Similarly, the coordinator can verify the signer's identity by validating the signature of the acknowledgment Data packet.
The derived secrets from DH protects subsequent data exchange.

\para{Confidentiality of the Signature Collection}
We analyze the confidentiality of the RPC protocol used in \sysname against eavesdroppers.
Specifically, we assume the eavesdropper can observe the Interest and Data packets exchanged between the coordinator and signers.
Among them, the potentially useful information includes \first packet names, \second key locator names, and \third the application parameters carried in the initial RPC request Interest.
The rest of the information is either protected by the encryption or does not leak useful information to the attacker (\eg, public DH parameters, signature values).
Here we iterate the revealed information from these sources.
\begin{itemize} [leftmargin=*]

\item Name of the RPC request (and acknowledgment):
The name contains the prefix of the signer and coordinator, thus revealing their identities.
Since the prefix is used for forwarding, it is inevitable unless additional infrastructures are used, \eg NDN-based VPN or onion routing.
The rest of the name components are either statically defined by the protocol or generated from the hash function of the payload, revealing no further information than the payload itself.

\item Name of the RPC parameter request (and reply):
Similarly, the only known information to the attacker is that the signer sends a request for the parameter; the rest of the name components are meaningless.

\item Name of the RPC result fetch request (and reply):
Since the packet name is generated irrelevantly to previous packets and cannot be predicted by the attacker, the unlinkability holds if there exists a sufficient anonymity set\footnote{A insufficient anonymity set results in easy compromise of the privacy. For example, if the coordinator only has contacted one signer in a given time period, it is easy to link a later Interest to the result fetch Interest.
We acknowledge this assumption is vague and not quantitative, but this is a general open issue for many privacy-enhancement systems.}.
That is, the attacker cannot recognize whether an Interest is for result fetching or not, so it cannot infer the RPC result.

\item Application parameters in the RPC request Interest:
The parameter packet name and DH parameters are included in the request Interest packet, but they do not leak useful information to the attacker; by knowing the parameter name, the attacker can only know whether the signer fetches the parameter, but cannot know if the signer will agree to sign the data or not.
The rest of the parameter, namely the DH parameters, can be considered random bits and thus makes no sense to the attacker.

\item Key locators:
In the exchange of the RPC request, key locators can reveal the identity of the signer and the coordinator, but as stated, they are already revealed by the names for forwarding purposes.
The rest of the key locators are used for HMAC, which is random and thus reveals no information about the packet signer.

\end{itemize}

Therefore, by observing the public information carried in the packets of RPC, an attacker cannot track the result of the signature collection process.
The key enabler is the unlinkability between an RPC and its result fetching process.

\para{Resistance to Record-and-replay Attack}
We consider a threat where an eavesdropper may record the signed RPC request Interest sent by the coordinator and attack a signer by replaying this Interest.
Such an attack can be effectively prevented by adding a timestamp and nonce into the Interest packet signature.
At the signer side, the signer can verify the timestamp of each incoming RPC request Interest against a predefined grace period.
In addition, the signer can keep a small state, say a bloom filter~\cite{BloomFilter}, of the nonce values seen from the recent legal requests.
In this way, the attacker cannot replay because there is no Interest packet whose timestamp is before the current time by the grace period and whose nonce has not been seen by the signer.

\para{Resistance to Denial-of-service Attack}
An attacker may try to deny the service of a signer by flooding RPC request Interests.
In general, the request Interest is signed and indicates the DH key generation, which consumes a certain amount of computation at the sender side.
In addition, these Interests cannot be copied from a recorded valid request, as stated before. These facts can discourage the attack.

However, the attacker may forge a large number of request Interests that are correct in the format but contain incorrect signature values or DH key bits.
Note that this is not a specific problem for \sysname but a general issue for many protocols, including TLS.
In NDN, such attacks can be mitigated by the DDoS defense mechanisms like FITT~\cite{Fitt}.

\section{Implementation and Evaluation}
\label{sec:evaluation}

We have implemented \sysname into an open-source library~\footnote{The codebase repository URL will be included with the final version of the paper} in C++.
Our library depends on ndn-cxx~\cite{ndn-cxx} and a BLS library~\cite{herumi-bls} that has been fully examined and is compatible with the BLS API specification defined by Ethereum 2.0~\cite{ethereum2-bls}.
The multisignature trust schema is implemented in the library as a trust schema configuration file parser.
To support more complicated use cases where multiple legitimate signer sets are valid, one can simply create multiple trust schema files applying to the same data prefix.
In the signature verification process, the signature is accepted if there is at least one trust schema rule that is satisfied.
The known signers are implemented with a separate configuration file that contains a list of NDN certificates.

In the rest of the section, we report the evaluation of our C++ implementation of \sysname.
Our evaluation result shows that
\first \sysname requires low-memory and low-network overhead to support multiparty signature signing, collection, and verification,
\second the cryptographic operations introduced by using \sysname is efficient and practical, and
\third \sysname is resilient to unavailable signers.

\para{Bandwidth and Network Overhead}
A BLS signature piece from a single signer and an aggregate signature is of the same size.
With BLS12-381 elliptic curve~\cite{pairing-friendly-curves-rfc} (providing about 128 bits of security), the BLS signature size is 96 bytes, which is slightly larger than an ECDSA signature (about 72 bytes) and much smaller than an RSA signature (256 bytes) while providing similar bits of security.
Regarding the size of public key certificates, a BLS public key over BLS12-381 is only 48 bytes, which is much smaller than public keys in ECDSA and RSA that provide a similar level of security.

The RPC in \sysname requires two(2) round trip times (RTTs) when the signer is available. Therefore, if the signer is unavailable, the coordinator will notice it after the timeout of the first Interest.

\para{NDN data Signing/Verification Overhead}
We compare the NDN data signing and verification performance of BLS multisignature with conventional public key signatures, including ECDSA (on elliptic curve secp256r1) and RSA (with 2048 bits key).
ECDSA and RSA signatures are realized by ndn-cxx's keychain with OpenSSL.
The content of the data to be signed is 16 bytes of random data.

\begin{table}[ht]
	\footnotesize
	\centering
	\begin{tabular}{ c | c c c}\hline
		\toprule
		Operation & BLS & ECDSA & RSA \\
		\midrule
		Signing & 1.38ms & 0.15ms & 1.58ms \\
		Verification & 4.32ms &  0.43ms  & 0.14ms \\
		\bottomrule
	\end{tabular}
%	\caption{Signing and Verification Time with Different Signature Schemes}
%	\label{tbl:sign-verify}
\end{table}

As shown, although slower than ECDSA and RSA (OpenSSL has specialized hardware and software optimization), the performance of BLS signature in NDN is still acceptable for practical use.

%\begin{figure}[t]
%	\centering
%	\includegraphics[width=0.49\textwidth]{figures/sign_types}
%	\caption{The execution time for different Signature Type\zhiyi{Need more discussion about what we want to show with this figure}}
%	\label{fig:sign_type}
%\end{figure}

\begin{figure} [t]
     \centering
     \begin{subfigure}[b]{0.235\textwidth}
         \centering
         \includegraphics[width=\textwidth]{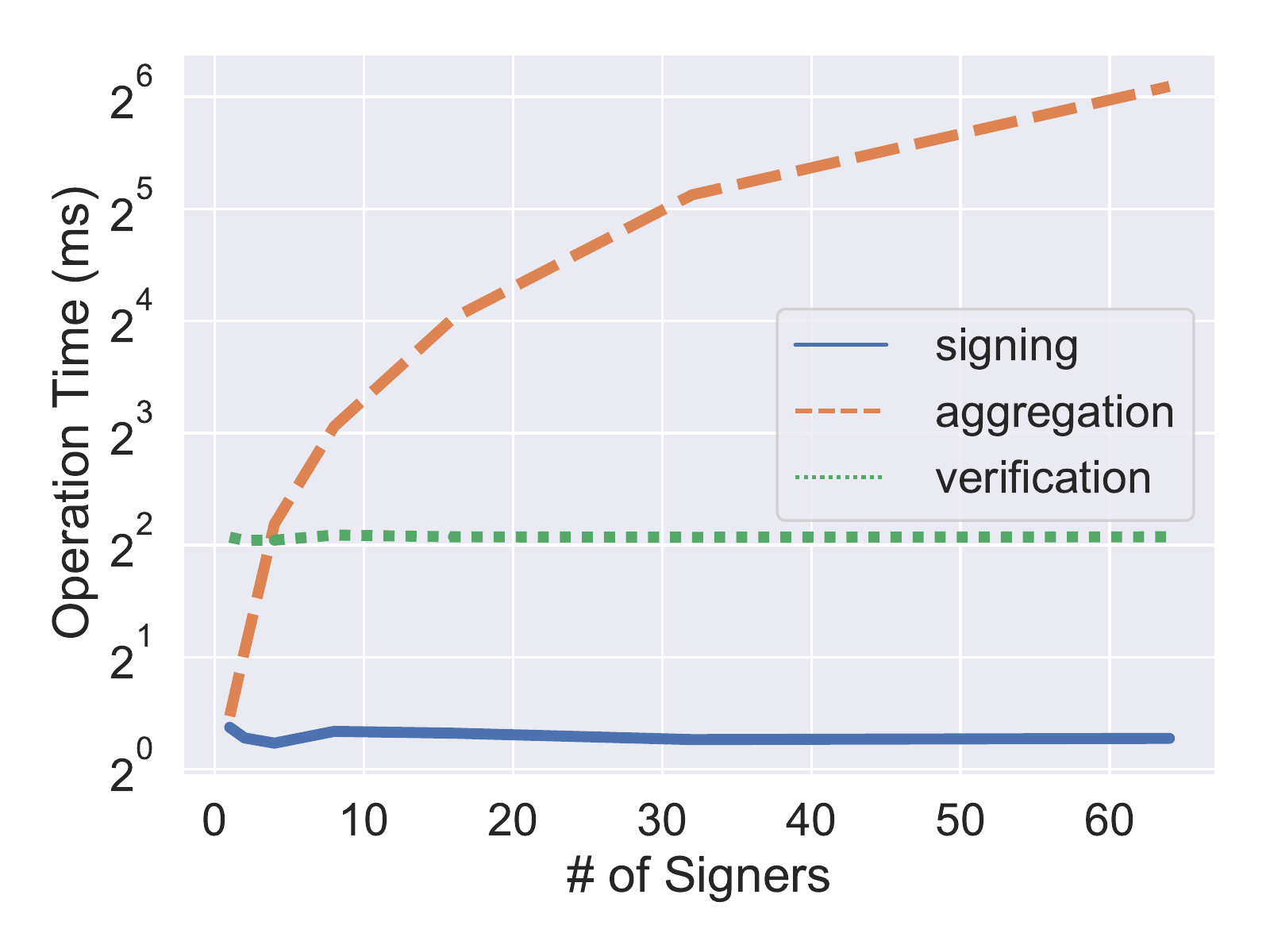}
         \caption{Different \# of Signers}
         \label{fig:signer_size}
     \end{subfigure}
     \hfill
     \begin{subfigure}[b]{0.235\textwidth}
         \centering
         \includegraphics[width=\textwidth]{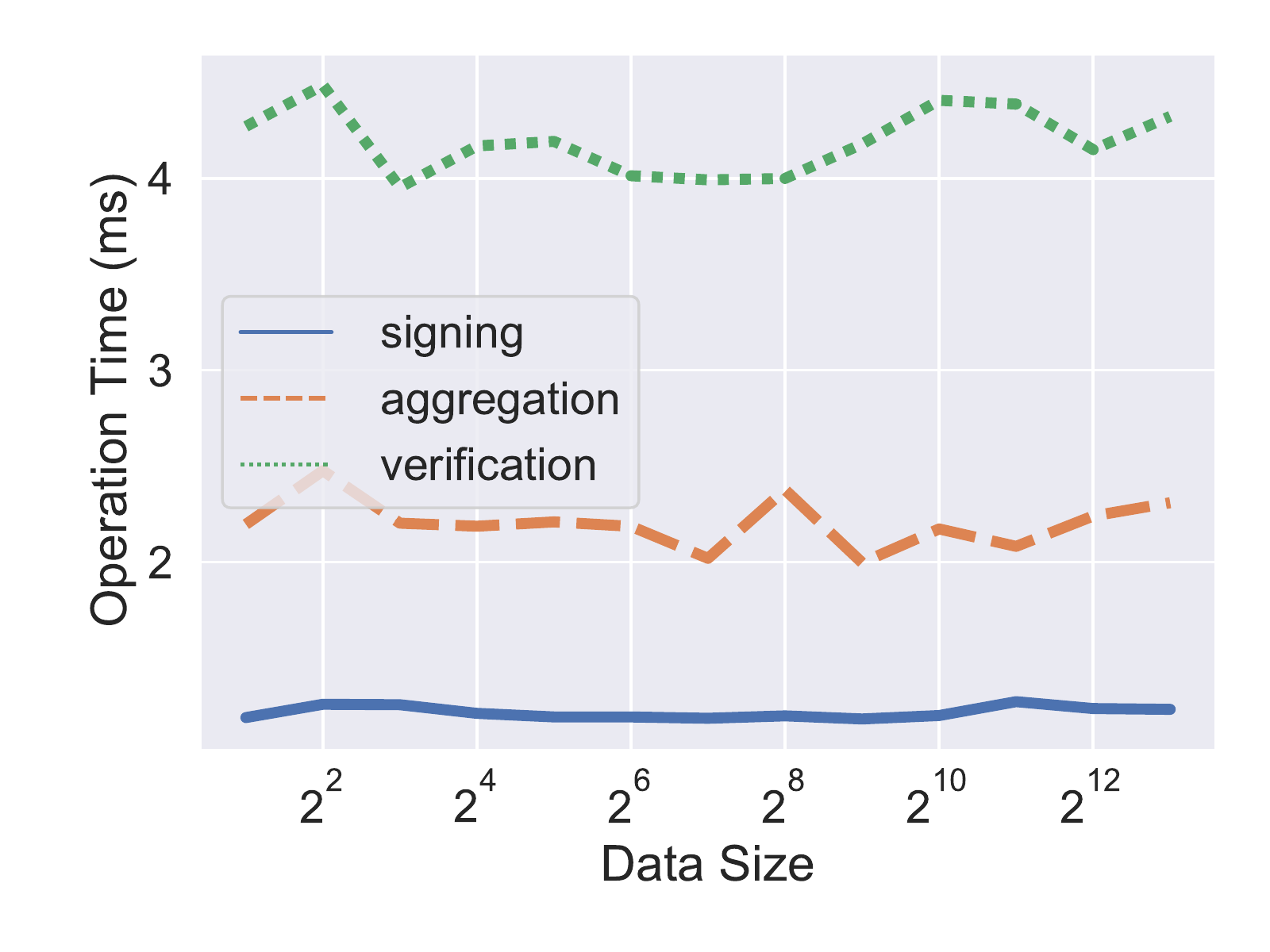}
         \caption{Different data Size}
         \label{fig:data_size}
     \end{subfigure}
     \vspace{-4mm}
     \caption{Scalability to Signer Number and data Size}

\end{figure}

\para{Scalability to Signer Number and Data Size}
We also evaluate the scalability of \sysname by increasing the number of signers needed for the signature generation and the size of data to be signed.
As shown in Figure~\ref{fig:signer_size}, only aggregation time is linearly affected by the signer number while the signing and verification time stays almost constant.
%he BLS signature time for
%The aggregation time is linear to the number of signers.
%The execution time is reported in
%The tests are done with 1000 trials, and the data content are 16 bytes of random data.
As reported in Figure~\ref{fig:data_size}, when increasing the data size with two signers, the time of signing, aggregation, and verification stay mostly constant.

%\begin{figure}[t]
%	\centering
%	\includegraphics[width=0.49\textwidth]{figures/signer_size}
%	\caption{The execution time for different signer size}
%	\label{fig:signer_size}
%\end{figure}

%\para{Scalability to the Size of the Data}
%
%The BLS signature time for signing, aggregation and verification stays mostly constant for different data size.
%The execution time is
%The tests are done with 1000 trials with 2 signers.

%\begin{figure}[t]
%	\centering
%	\includegraphics[width=0.49\textwidth]{figures/data_size}
%	\caption{The execution time for different data content size}
%	\label{fig:data_size}
%\end{figure}
%\section{Related Work}
%
%\begin{itemize}
%	\item BLS based systems and other multiparty trust systems in TCP/IP
%	\item Existing NDN based multiparty trust system
%	\item Secret Sharing: Why use system instead of crypto for 3-out-of-5. Trade-offs discussion.
%\end{itemize}
\section{Discussion}
\label{sec:disc}

\subsection{Data Object To Be Signed}

\sysname helps to sign a data object which can mainly be in two different forms.
If the data object can fit into a single NDN Data packet~\footnote{A data object can also be an Interest packet, then the result will be a signed Interest~\cite{signedInterest}.}, it can be directly embedded in the RPC parameter packet.
A data object can also be larger than the NDN maximum packet size.
In this case, the object cannot be directly embedded into one RPC parameter packet, \eg, a large file or stream data.
Instead, it needs to be segmented into multiple packets.
To avoid multiple signers from signing each segment, we can introduce a manifest Data packet~\cite{ndn-manifest}, which contains the hash values of all segments in the form of a Merkle Tree and their names.
Then, only the manifest Data needs to be embedded in the RPC parameter packet and signed by multiple parties.

\subsection{Design Discussion of \sysname RPC}

We discuss some of the design alternatives in \sysname RPC and explain why these alternatives were not adopted.

\myquestion{Which packet should carry the parameters}
In \sysname, the initial request Interest to the signer contains the name of the unsigned data instead of the data itself for the following reasons.
First, in the initial request Interest packet, the coordinator has not finished the Diffie-Hellman exchange and thus does not know how to encrypt the unsigned data.
Second, directly carrying the unsigned data does not scale well when the data is large.

%The actual unsigned data need to be fetched by the signer again.
%We choose the design to alleviate the network overhead at the routers.
%In NDN, each router will cache the interest for Interest Aggregation.
%Therefore, having the data directly in the Interest will take up a large space in each router on the forwarding path.
%Furthermore, placing the data in a data wrapper would allow the different signer to fetch the same piece of data,
%which can improve the performance by exploiting the in-network caching in NDN.
%

\myquestion{How to fetch the result}
\sysname uses separate Interest packets for the request and the result.
An alternative design would be to let the signer directly reply with the RPC result to the request Interest (\ie, the first Interest packet sent by the coordinator in the RPC).
However, this does not work when the signer's processing time is longer than the Interest timeout.
Given that the coordinator needs to determine the signer's availability quickly, the coordinator cannot set the timeout of the request Interest to be too long.
%In addition, the timeout of the request Interest cannot be set too long because this timeout is importantly for the coordinator to be aware of the availability of the signer.

\myquestion{How to publish the result}
In \sysname, the RPC result is encapsulated in a result Data packet.
An alternative method would be to directly publish the signed data as a Data packet for fetching; however, this design may result in several issues.
For one, this approach may result in cache poisoning because the result Data packet shares the same name as the Data packet carrying the aggregate signature.
Additionally, directly publishing the result data can reveal the signed content and disclose that the signer has signed the data, which can break the confidentiality of the signing process.
Even though the data content can be encrypted, the data name still cannot be hidden for the data retrieval purpose.
%We design these pieces of names to ensure each data have a unique name.
%This is fundamental in NDN, because it allows for in-network caching of data.
%If we did not disambiguate the data, the incorrect version of data could be fetched from the network.

%\subsection{Multisignature without A Coordinator}
%
%xxx
\section{Conclusion}
\label{sec:conclusion}

%%To the best of our knowledge,
\sysname is an application-independent multisignature toolset designed to support and automate multiparty signature signing, collection, and verification.
Through our work, we show how \sysname leverages existing NDN features
to support more complex trust models than the conventional producer-consumer model.
%% and thus suitable for different application scenarios.
On the other hand, the security support of today's TCP/IP network, \eg, the TLS, is limited to the point-to-point trust model.
Thus, new security models or cryptographic tools such as multisignature are mainly supported at the application layer, separate and orthogonal to the overall network security architecture.
In contrast, NDN's data-centric security based on application layer semantic naming enables a coherent overall system security framework, which poses no limitations to new trust models.
It provides a simple yet graceful platform to explore new security models and tools with coherent network application support.

In our future work, we will continue to exercise \sysname in the DER scenarios.
In addition, we will also look into and understand new security use cases and develop new solutions for NDN security support.

%
%
%With its flexibility, \sysname are suitable for further improvements and use cases.
%Future research will investigate
%\first how to select the potential signers effectively,
%\second scalability with a complex schema,
%and \third integration with message transport applications.

\bibliographystyle{acm}
\bibliography{paper}

\end{document}